\documentstyle{article}

\title{\bf On the jet spectrum
in nucleus-nucleus interactions}
\author{M.A.Braun$^1$, E.G.Ferreiro$^2$, C.Pajares$^2$ and D.Treleani$^3$ \\
 $^1$ Department of High
Energy Physics,
University of St. Petersburg,\\ 198904 St. Petersburg, Russia\\
 $^2$ Departamento de Fisica de Particulas, Unicersidade de Santiago
 de Compostela\\
15706 Santiago de Compostela, Spain\\
 $^3$ Dipartimento
di Fisica Teorica dell'Universit\`a di Trieste and INFN, Sezione di Trieste\\
Strada Costiera 11, Miramare-Grignano,
      I-34014, Trieste, Italy    }

\date{}
\def\beq{\begin{equation}}
\def\eeq{\end{equation}}

\begin{document}

\maketitle
\vspace{1 cm}
\begin{abstract}
\noindent
We derive the inclusive transverse spectrum of minijets in nuclear collisions at
very high energies, by assuming that the nuclear $S$-matrix
factorizes as a product elastic
$S$-matrices of elementary partonic collisions.
Interference effects and, in particular, the
contribution of loop diagrams are fully taken into account in the derivation of the spectrum,
which is shown to coincide with the result already obtained by superposing
the elementary interactions incoherently. A quantitative
analysis confirms that the deformation induced by multiple collsions
is a large effect at RHIC and LHC energies, for transverse momenta $\sim$20 GeV.
\end{abstract}
\vspace{1 cm}


\section{Introduction}
The growing experimental activity in heavy-ion collision has stimulated
a renewed interest on the $p_T$ distributions of
secondaries in AB collisions. From the theoretical point of view a
special attractiveness presents the production of jets, which, in
principle, admits a study in the framework of perturbative QCD. In
fact theoretical predictions for the inclusive jet production in
$p\bar p$ collisions  at 1800 GeV agree quite well with the
experimental data for jet transverse energies in the region
50$\div$ 250 GeV/c\cite{Affolder:2001fa}. In
the region of lower jet transverse momenta,
the agreement is not as good and phenomenogical parameters need to be
introduced\cite{Eskola:1988yh}\cite{Eskola:2001rx}. In nuclear collisions the situation becomes
obviously much more complicated. There are several mechanisms
which influence the form of the jet spectrum and make it different
from the case of $p\bar p$. In the nucleus the partonic
distributions are changed, the produced jets may loose some energy
while passing through the thick nuclear matter, jets may
accumulate more transverse momentum by hard rescatterings. All
these different phenomena need to be disentangled to gain a
comprehensive understanding of the production mechanism.

In the present paper we focus on the effects of rescattering on
the transverse spectrum of minijets in nucleus-nucleus collisions.
The problem was studied in the past assuming that the total
inelastic hard cross-section can be obtained from  the forward
scattering amplitude by substituting the partonic (elastic)
cross-sections for the forward parton-parton scattering
amplitude\cite{Calucci:1989hb}\cite{Calucci:1991sz}\cite{Accardi:2001ih},
similarly to the analogous case, where the
inelastic nucleus-nucleus cross section is expressed in terms of
the nucleon-nucleon total cross-sections
\cite{Bialas:1976ed, Pajares:1985}. As a
consequence, in the approach of Refs.
\cite{Calucci:1989hb}\cite{Calucci:1991sz}\cite{Accardi:2001ih}, the hard
nucleus-nucleus cross section is expressed as an incoherent
superposition of interactions between the partons of the nuclear
structures. Possible coherence effects are not included and, in
particular, the well known problem of loops, arising in the
Glauber approach to nucleus-nucleus collisions\cite{Boreskov:yt}, is completely
disregarded.

The purpose of the present paper is to take interferences
into account and
include all loops in the evaluation of rescatterings in the
inclusive spectrum of minijet production in nucleus-nucleus
collisions. In our approach each colliding nucleus will be assumed
to consist of a variable number of partons, distributed both in
impact parameter and rapidity, while the nuclear $S$-matrix will
be assumed to factorize as a product of parton-parton
$S$-matrices. The inclusive cross-section for minijet production
will be derived without approximations, apart from the usual
disentangling of the longitudinal and transverse degrees of
freedom and, as we will show, the result will turn out to be
identical to the expression already obtained in
\cite{Calucci:1989hb}\cite{Calucci:1991sz}\cite{Accardi:2001ih} under the
incoherent no-loop approximation.

As an application, in the second part of the paper we study the
$p_T$ distributions in central Pb-Pb collisions at nucleon-nucleon c.m energies of 200
and 6000 GeV, corresponding to the regimes at RHIC and LHC.
Since our aim is not to fit the existing data but rather to see
the influence of rescatterings, we compare our results
with the naive picture where each projectile parton is allowed to interact
with a single target parton only, weighted with the
partonic distributions of both nuclei, which are taken to be
identical to those of the proton.  Rescattering effects
are obviously maximal at relatively small transverse momenta, where the
partonic cross-section is relatively large. For this reason we
concentrate our attention on the production of minijets with
transverse momenta up to 10 $\div$ 20 GeV/c. To make our picture
compatible with the existing experimental data in $p\bar p$
collisions at this momenta, we shall use an efective partonic
cross-sections taken from the lowest order pertubative QCD and
corrected by phenomenological parameters as in \cite{Eskola:1988yh}\cite{Eskola:2001rx}.

\section{Glauber theory for hard parton scattering}
We start by introducing the expression for the hard scattering
amplitude in terms of partonic interactions by following the philosophy of the Glauber
approach to nucleus-nucleus collisions. Hence we assume
1) conservation of partonic  longitudinal momenta
and impact parameters and 2)  factorization of the  $S$ matrix into the
product of elementary partonic $S$ matrices.

The nucleus state $|A\rangle$ can be represented by a superposition
of states with a different number $n$ of partons, each characterized by its
scaling variable $x$ and impact parameter $b$. For brevity we denote
$z=\{x,b\}$. We also consider only one partonic flavour, although the generalization
to
different flavours $f$ is trivial.
With these notations one has
\beq
|A\rangle=\sum_{n}\int\prod_{i=1}^nd^3z_i\Psi_{A,n}(z_i)|n,z_i\rangle
\eeq
In the following we shall denote the phase space of the $n$-parton
configuration in nucleus A as
\beq
d\tau_A(n)=\prod_{i=1}^nd^3z_i
\eeq
For the nucleus A - nucleus B collision the $S$ matrix has the form
\[
\langle A'B'|S|AB\rangle= \sum_{n,l,n',l'}\int
d\tau_A(n)d\tau_A(n')d\tau_B(l)d\tau_B(l')\]\beq
\Psi_{A',n'}^*(z'_i)\Psi_{B',l'}^*(u'_j)
\Psi_{A,n}(z_i)\Psi_{B,l}(u_j) \langle
n',z'_i;l',u'_j|S|n,z_i;l,u_j\rangle \eeq
Here $u_j$ are the
partonic variables of the nucleus B, being $l$ the number of partons in a
given configuration. As stated, our first assumption is that
the $S$ matrix is diagonal in the basis $\{n,z_i;l,u_j\}$:
\beq
\langle n',z'_i;l',u'_j|S|n,z_i;l,u_j\rangle= \delta_{nn'}
\delta_{ll'}\prod_{i=1}^n\delta^3(z_i-z'_i)
\prod_{j=1}^l\delta^3(u_i-u'_i)S_{nl}(z_1,...z_n|u_1,...u_l) \eeq
Physically this assumption corresponds taking into account binary collisions only,
therefore conserving the number of partons in the hard process, which may be justified
arguing that in a hard interaction parton
production is damped by an extra power of the strong coupling
constant. We also assume that neither the longitudinal momentum
nor the impact parameter are changed in a hard collision, which are
basic assumptions in the Glauber approach. The range in $x$ and $p_{\perp}$
for which the treatment
can be applied is therefore restricted to the region where
longidudinal momenta are much higher as compared to the transferred momenta.
Under assumption (4) the $S$ matrix simplifies to
\beq \langle A'B'|S|AB\rangle= \sum_{n,l}\int d\tau_A(n)d\tau_B(l)
|\Psi_{A,n}(z_i)|^2|\Psi_{B,l}(u_j)|^2
S_{nl}(z_1,...z_n|u_1,...u_l) \eeq

A second assumption intrinsic in the Glauber approach is that of factorization
of the $S$ matrix in (5)
into a product of elementary $S$ matrices for parton-parton scatterings
\beq
S_{nl}(z_1,...z_n|u_1,...u_l)=\prod_{i=1}^n\prod_{j=1}^ls_{ij}
\eeq
where $s_{ij}$ is the $S$ matrix for the collision of parton $i$ from
nucleus A with parton $j$ from nucleus B:
\beq
s_{ij}=1+ia(z_i,u_j)\equiv1+ia_{ij}
\eeq
$a$ being the parton-parton scattering amplitude.
Since we assume that no parton production takes place, $s_{ij}$
corresponds to purely elastic scatterings (or to binary reactions in the
multi-flavour case). Unitarity is therefore satisfied in its simplest form
\beq
(s_{ij})^*s_{ij}=1
\eeq
(in the multi-flavour case this is changed into a matrix condition in the
flavour space)

To complete the picture we have to introduce the normalization
condition for the nuclear wave function, which
reads
\beq \sum_{n}\int d\tau_A(n)|\Psi_{A,n}(z_i)|^2=1
\eeq
As one observes the normalization condition does not limit
the parton population of a given nuclear configuration,
represented by a given term in the sum, it only restricts the
total probability of the different configurations to be unity. The
cross-section introduced in [4] - [6] corresponds to choosing the
distribution in the number of partons to be Poissonian:
\beq
w_n=\int d\tau(n)|\Psi_{A,n}(z_i)|^2=e^{-\langle n\rangle}
\frac{\langle n\rangle^n}{n!} \eeq
If one additionally assumes
that the the nuclear partonic wave function factorizes as
\beq
|\Psi_{A,n}(z_i)|^2=c_n\prod_{i=1}^n\Gamma(z_i) \eeq
one immediately gets
\beq c_n=\frac{1}{n!}e^{-\langle
n\rangle},\ \ \langle n\rangle=\int d^3z\Gamma(z) \eeq

With the forward AB scattering amplitude defined by (5)- (7), we can
proceed to determine the relevant cross-sections. The total cross-section
is of course obtained from the imaginary part of the forward
$\langle AB|S-1|AB\rangle$. However we are not interested in the total
cross-section but rather in the total hard cross-section with large
momenta of observed partons. In the diagrammatic language this cross-section
corresponds to cutting the forward scattering amplitude through the
intermediate partonic propagators, with the restriction that the cut partons
should have a large transverse momentum component.

One can standardly present all contributions to (5) - (7) in terms of
diagrams which show partonic interactions via the amplitudes $ia_{ij}$.
External partons correspond to those present in the colliding nuclei.
They are assumed to have small transverse momenta and to be distributed in
$x$ and $b$ according to $|\Psi_{A(B),n}|^2$. All diagrams can be separated
into tree diagrams and diagrams with loops. In the tree diagrams all
transferred momenta are small ("zero"), so that they are expressed through
the forward scattering amplitudes $ia_{ij}$. Loop diagrams involve
integrations over the intermediate parton momenta, which may be large.
So they cannot be expressed via the forward scattering amplitudes $ia_{ij}$
but rather involve these amplitudes at large transferred momenta.

If one restricts oneself to the tree diagrams then the only way to find
intermediate partons with high transverse momenta is to cut the forward
scattering amplitudes $ia_{ij}$ converting them into the total (elastic)
partonic cross-section $\sigma^{el}$. According to the AGK rules,
the sum over all possible cuts gives the final hard cross-section in the form
$\langle AB|S-1|AB\rangle$ with $ia_{ij}$ substituted by $-\sigma^{el}_{ij}$
in (7). This latter formula was the starting point in the derivations in
\cite{Calucci:1989hb}\cite{Calucci:1991sz}\cite{Accardi:2001ih}.

However loops give new contributions to the hard cross-sections,
since cutting the loop produces intermediate partons with large
transverse momenta. The result cannot be expressed in any simple
way via the partonic cross-sections $\sigma^{el}$, since it
involves partonic amplitudes at large momentum transfers.
Unfortunately it is easy to check that, at least perturbatively,
the loop contributions are dominant. The reason is that the
partonic amplitude $a$ is predominatly real (of order
$\alpha_s$, whereas its imaginary part (and $\sigma^{el}$) is of
the order $\alpha_s^2$ and thus supressed by a factor $\alpha_s$). As an explicit example,
let us compare the different contributions shown in Fig. 1. The external
lines of the interacting partons supply a factor $A^{1/3}$ or
$B^{1/3}$ each, due to the partonic densities in the nuclei. By
cutting the single scattering diagram $a$ one obtains a
contribution of order $(AB)^{1/3}\alpha^2_s$, while its iteration
$b$ gives a contribution of order $(AB)^{2/3}\alpha_s^4$. The contributions
of the cuts of the
Glauber-like diagrams $c$ and $d$ are of orders
$A^{1/3}B^{2/3}\alpha_s^4$ and $A^{2/3}B^{1/3}\alpha_s^4$. Tree
diagrams with 3 interactions, as diagram $e$ in the figure,
give a contribution of
order $(AB)^{2/3}\alpha_s^6$. The leading contribution of the loop diagram
$f$ is, on the contrary of order $(AB)^{2/3}\alpha^4$.
Since all 4 elementary amplitudes can now be taken uncut (and real) in that case,
this last
contribution is by far dominant, as compared to all connected diagrams,
and comparable to the iterated lowest order diagram. It can be
shown that a similar argument holds for any number of interacting partons: the
dominant contribution is given by iterations of the sum of the
lowest order diagrams plus loops. All other diagrams are suppressed by
a factor $\alpha_s^2$.

As a consequence by substituting $ia$ by $-\sigma^{el}$, which is
the approximation used in \cite{Calucci:1989hb}\cite{Calucci:1991sz}\cite{Accardi:2001ih},
one only selects a limited,
sub-leading, set diagrams, the tree diagrams. Notice that the
situation is quite different as compared to the case of the
Glauber picture of AB scattering in terms of nucleons. In that
case the amplitude is in fact mainly imaginary and, in addition,
loops are likely to be suppressed by finite formation time
arguments. The inclusive hard cross-section, obtained in \cite{Calucci:1989hb}\cite{Calucci:1991sz}\cite{Accardi:2001ih} in
the tree diagram approximation, needs therefore to be re-derived
keeping the contribution of loops into account.

\section{Jet spectrum: general}
To find the spectrum of emitted jets we have to fix the final state
of one of the nuclei in a specific manner, in which one of the partons
appears with the observed wave function $\psi_{\alpha}(z)$.
The choice of the observed parton
is irrelevant, since the total wave
function of the nucleus is supposed to be symmetric in all the partons, so
we choose it to be the first one. Note that choosing the observed parton
to belong to nucleus A implies studying the spectrum in the direction of
nucleus A and conversely for nucleus B. In the following we study the former case, so we take
\beq
\Psi_{A',n}(z_1,z_2,...z_n)=
\sqrt{n}\psi(z_1)_{\alpha}\tilde{\Psi}_{A',n-1}(z_2,...z_n)
\eeq
The second factor represents the state of the remaining $n-1$ partons,
together with the observed one. The factor $\sqrt{n}$ is to adjust the
normalizations of the symmetrized wave functions $\Psi_n$ and
$\tilde{\Psi}_{n-1}$.
Using (13) we get the probability to observe a parton with wave
function
$\psi$ at a given impact parameter $\beta$ in the nucleus A-nucleus B
scattering as
\[
\frac {d\sigma_{\alpha}}{d^2\beta}=
\sum_{nl,n',l'}\sqrt{nn'}\sum_{A',B'}\int
d\tau'_A(n')d\tau'_B(l')
d\tau_A(n)d\tau_B(l)\]\[
\Psi_{A,n'}^*(z_i')\Psi_{B,l'}^*(u_i')
\Big[S^*_{n'l'}(z'_i|u'_j)-1\Big]\psi_{\alpha}(z_1')
\tilde{\Psi}_{A',n'-1}(z'_2,..z_n')\Psi_{B,l'}(u'_j)\]\beq
\psi^*_{\alpha}(z_1)\tilde{\Psi}_{A,n}^*(z_2,...z_n)
\Psi_{B',l}^*(u_j)\Big[S_{nl}(z_i|u_j)-1\Big]
\Psi_{A,n}(z_i)\Psi_{B,l}(u_j)
\eeq
The sum over all intermediate partonic states, with a given number of
partons, is done by using closure:
\beq
\sum_{A}\Psi_{A,n'}(z'_1,...z'_{n'})\Psi_{A,n}^*(z_1,...z_{n})
=\delta_{nn'}\prod_{i=1}\delta(z_1',..,z_n|z_1,...,z_n)
\eeq
where the $\delta$ function is a symmetrized product of
$\delta^3(z'_i-z_i)$.
We obtain
\[
\frac {d\sigma_{\alpha}}{d^2\beta}=
\sum_{nl}n\int dz_1dz'_1\psi_{\alpha}(z_1')\psi^*_{\alpha}(z_1)
d\tau_A(n-1)d\tau_B(l){\Psi_{A,n}}^*(z_1',z_2,...z_n)
\Psi_{A,n}(z_1,z_2,...z_n)|\Psi_{B,l}(u_j)|^2\]\beq
\Big[S^*_{nl}(z'_1,z_2,...z_n|u_1,...,u_l)-1\Big]
\Big[S_{nl}(z_1,z_2,...z_n|u_1,...u_l)-1\Big]
\eeq

Notice that by summing over all possible observable states $\psi_{\alpha}$
one gets
\[\sum_{\alpha}\psi_{\alpha}(z_1')\psi_{\alpha^*}(z_1)=\delta(z_1'|z_1)\]
which implies
\[
\sum_{\alpha}\frac {d\sigma_{\alpha}}{d^2\beta}=
\sum_{nl}n\int
d\tau_A(n)d\tau_B(l)|\Psi_{A,n}|^2(z_i)
|\Psi_{B,l}(u_j)|^2\]\[
\Big[S^*_{nl}(z_1,z_2,...z_n|u_1,...,u_l)-1\Big]
\Big[S_{nl}(z_1,z_2,...z_n|u_1,...u_l)-1\Big]\]\[=
\sum_{nl}n\int
d\tau_A(n)d\tau_B(l)|\Psi_{A,n}|^2(z_i)
|\Psi_{B,l}(u_j)|^2\]\beq
\Big[2-S^*_{nl}(z_1,z_2,...z_n|u_1,...,u_l)
-S_{nl}(z_1,z_2,...z_n|u_1,...u_l)\Big]=
<n>\sigma_{AB}^{tot}
\eeq
The result is expected: the integrated inclusive cross-section gives
the total cross-section multiplied by the average multiplicity, which in our case
is obviously $<n>$. Since $<n>$ is supposed to grow with $A$
only linearly (or even less rapidly, if the EMC effect is taken into account)
the multiplicity does not grow faster then $A^1$.

\section{Jet spectrum: non-trivial and geometric contributions}
The product of $S$ matrices in (16) generates 4 terms which we
rearrange as
\[
\Big[S^*_{nl}(z'_1,z_2,...z_n|u_1,...,u_l)
S_{nl}(z_1,z_2,...z_n|u_1,...u_l)-1\Big]\]\beq
-\Big[S^*_{nl}(z'_1,z_2,...z_n|u_1,...,u_l)-1\Big]
-\Big[S_{nl}(z_1,z_2,...z_n|u_1,...u_l)-1\Big]
\eeq
It is instructive to study the meaning of the three terms from the
point of view of the intermediate physical partonic states,
corresponding to cutting the whole diagram with partonic interactions
in the overall unitarity relation. In the second  term
only conjugate  amplitudes $[ia_{ij}]^*$ enter, which
means that this contribution corresponds to cut
the interaction diagram from the extreme right, namely to the right of all
partonic interactions. Conversely in the last term only the amplitudes $a_{ij}$ enter,
which corresponds
to cut the amplitude to the left of all partonic
integractions. In both cases the momenta of the intermediate partons
are convoluted with their initial distributions in the
colliding nuclei, which is
very narrow for geometrical reasons, so that also the resulting spectrum cannot be much
broader.

To be more explicit, consider the third term in (18). Take the contribution from a
single interaction of the first
parton at impact parameter $b_1$ with some parton in B with its impact
parameter $c$. Since
 we are interested in the distributions
of partons from nucleus A, as a function of the transverse momentum and at a fixed rapidity,
we take
\beq
\psi_{\alpha}(z_1)=e^{ipb_1}\delta(x_1-x)
\eeq
where $x$ and $p$ are the scaling variable and transverse momentum
of the observed parton. Then the integration over $b$ and $c$ will take the form
\beq
\int d^2b_1d^2c e^{ipb_1}\Psi_{A,n}(b_1,...)ia(b_1-c)|\Psi_{B,n}(c,..)|^2
\eeq
where we have suppressed all the variables, with the only exception of $b_1$ and $c$. The typical scale
of the partonic interaction is much smaller than the nuclear scale,
so to a good approximation one can rewrite (20) as the product
\beq
\int d^2b_1 e^{ipb_1}\Psi_{A,n}(b_1,...)|\Psi_{B,n}(b_1,..)|^2
\int d^2c ia(c)
\eeq
It is evident now that this expression is different from zero only for values
of $p$ of the order of the inverse of the nuclear radius, unless one assumes that
the nuclear distributions itself contains partons with high transverse momenta.

The non-trivial part of the partonic distribution originated by the hard
collisions corresponds therefore to cutting the interaction diagram in between
hard collisions and thus is totally contained in the first term in (18).
Remarkably this contribution can be greatly simplified (differently with respect to
the last two terms in (18)).
In fact, using the factorization property (6)
 we can write
\beq
S_{nl}(z_1,z_2,...|u_1,...u_l)=
\prod_{j=1}^ls_{1j}\prod_{i=2}^n\prod_{j=1}^ls_{ij}
\eeq
Due to the unitarity relation (8) the second product in (22) is cancelled by
its conjugate in the product $S^*S$ in (18), so that we get
\beq
S^*_{nl}(z'_1,z_2,...z_n|u_1,...,u_l)
S_{nl}(z_1,z_2,...z_n|u_1,...u_l)-1=
\prod_{j=1}^l[1+ia(z'_1,u_j)]^*[1+ia(z_1,u_j)]-1
\eeq

Putting this expression into (16) we note that (23) does not depend on the
partonic variables $z_2,...z_n$ of nucleus A. Integrating over $z_2,...z_n$
and summing over $n$ one obtains the $\rho$ matrix of nucleus A as a function of the
nuclear partonic degrees of freedom:
\beq
\sum_n n\int d\tau_A(n-1)
\Psi_{A,n}^*(z_1',z_2,...z_n)
\Psi_{A,n}(z_1,z_2,...z_n)=\rho_A(z_1|z'_1)
\eeq
This relation in fact defines the nuclear $\rho$-matrix. As a consistency check one may notice
that, under the factorization assumption (11), at $z'_1=z_1$, (24) reduces
to $\Gamma(z_1)$.

The contribution from the 1st term in (18) may therefore be written as
\beq
\frac {d\sigma_{\alpha}^{(1)}}{d^2\beta}=
\sum_{l}\int dz_1dz'_1\psi_{\alpha}(z_1')\psi^*_{\alpha}(z_1)\rho_A(z_1|z'_1)
d\tau_B(l)|\Psi_{B,l}(u_j)|^2
\Big\{\prod_{j=1}^l[1+ia(z'_1,u_j)]^*[1+ia(z_1,u_j)]-1\Big\}
\eeq

\section{The hard spectrum}
To obtain a more explicit form of the spectrum
we make use of the factorization ansatz (11) for nucleus B:
\beq
|\Psi_{B,l}(u_j)|^2=\frac{1}{l!}e^{-\langle l\rangle}
\prod_{j=1}^l\Gamma(u_i)
\eeq
with
\beq
\langle l\rangle=\int d^3u\Gamma_B(u)
\eeq
We also assume for the parton distributions the factorized expression
\beq
\Gamma_B(u)=T_B(c) P_B(w)
\eeq
where $w$ and $c$ are the fractional momentum and transverse parton coordinate respectively.
The normalizations are
\beq
\int d^2bT_B(c)=1,\ \ \int dwP_B(w)=\langle l\rangle
\eeq
Expressing the wave function of the observed parton as in Eq.(19), we obtain
\[
(2\pi)^2\frac {d\sigma^{(1)}}{d^2\beta dy  d^2p}=
\sum_{l}\frac{1}{l!}e^{-\langle l\rangle}
\int d^2b_1d^2b'_1e^{ip(b_1-b'_1)}\rho_A(x,b_1-\beta|x,b'_1-\beta)
\]\beq
\prod_{j=1}^ld^2c_jdw_jT_B(c_j)P_B(w_j)
\Big\{\prod_{j=1}^l[1+ia(z'_1,u_j)]^*[1+ia(z_1,u_j)]-1\Big\}
\eeq
with $z=(x,b)$ and $z'=(x,b')$.
Taking the origin in the center of nucleus B, so that the
partonic distribution in nucleus A becomes shifted by the overall impact
parameter $\beta$,
due to factorization, the integrations over $(w_j,c_j)$
give the $l$-th power of
\beq
J=\int d^2c dwT_B(c)P_B(w)[1+ia(x,w;b'_1-c)]^*[1+ia(x,w,b_1-c)]
\eeq
where the dependence of the scattering amplitude on the
transverse distance between the interacting partons is explicit. The different terms
in this integral are treated as in the standard Glauber derivation.
Let us consider the term with the product of two amplitudes. The distance
between the interacting partons
 for hard interactions is very small as compared to the nuclear distances. So
using $r=b_1-c$ as a integration variable we find $c=b_1-r$ and
we can take the nuclear profile function out of the integral at
$c\simeq b_1$. We obtain
\beq
T_B(b_1)\int d^2rdwP_B(w)a^*(x,w;b'_1-b_1+r)a(x,w;r)\equiv
T_B(b_1)F_B(x,b'_1-b)
\eeq
where
\beq
F_B(x,b)=\int d^2rdwP_B(w)a^*(x,w;b+r)a(x,w;r)
=\int dwP_B(w)\int\frac{d^2p}{(2\pi)^2}I(p)e^{ipb}
\eeq
is the Fourier transform of the transverse momentum distribution
in the parton-parton interaction, averaged over the parton rapidity
distribution of nucleus B.
Note that at $b=0$ (33) gives the total cross-section for parton-parton
collisions (which coincides with the elastic cross-section), averaged
over the rapidities of the B-partons
\beq
\sigma^{tot}_B(x)=\sigma^{el}_B(x)=
\int dwP_B(w)\int d^2r |a(x,w;r)|^2
\eeq
The other terms in (31) are evaluated in a similar manner, the result is
\beq
J=\langle l\rangle+T_B(b_1)F_B(x,b'_1-b)-T_BF_B(x,0)
\eeq
Taking the $l$-th power of (35) and summing over all values of $l$ one gets
\[
\exp \Big( T_B(b_1)[F_B(x,b'_1-b_1)-F_B(x,0)]\Big)
\]
The term in (30) with unity gives unity after summation over $l$, so that the cross section
is expressed as
\[
(2\pi)^2\frac {d\sigma^{(1)}}{d^2\beta dy d^2p}=
\int d^2b_1d^2b'_1e^{ip(b_1-b'_1)}\rho_A(x,b_1-\beta|x,b'_1-\beta)
\]\beq
\Big\{e^{ T_B(b_1)[F_B(x,b'_1-b_1)-F_B(x,0)]}
-1\Big\}
\eeq
Introducing the integration variables $r=b'_1-b_1$ and $b=(1/2)(b'_1+b_1)$ and
keeping into account that $r$ is a small quantity on the nuclear scale, one obtains
\beq
(2\pi)^2\frac {d\sigma}{d^2\beta dy d^2p}=
\int d^2bd^2re^{ipr}\rho_A(x,b-\beta)
\Big\{e^{ T_B(b)[F_B(x,r)-F_B(x,0)]}
-1\Big\}
\eeq
Here $\rho_A(x,b)=T_A(b)P_A(x)$ is just the partonic distribution
of nucleus A.
Since $F_B(x,r)$ goes to zero as $r\to\infty$
the inclusive cross-section (37)
contains also a soft component of the spectrum, which in the expression
is represented by the term proportional to $\delta^2(p)$.
To remove it we subtract from the integrand its value at $r=\infty$.
Our final expression for the inclusive
cross-section is therefore
\beq
(2\pi)^2\frac {d\sigma}{d^2\beta dy d^2p}=
\int d^2bd^2re^{ipr}T_A(b-\beta)P_A(x)
\Big\{e^{ T_B(b)[F_B(x,r)-F_B(x,0)]}
-e^{-T_B(b)F_B(x,0)}\Big\}
\eeq

The only trace in (38) of the nucleus A is in the term
$T_A(,b-\beta)P_A(x)$,
which appears as a weight factor for the different contributions from the various
parts of nucleus B. For central collisions one can separate it as
a nearly constant factor, so that the form of the spectrum should be
practically independent of A and, in particular, it coincides with the spectrum
in hadron-nucleus B collisions.

As mentioned in the Introduction, this formula coincides with the
one obtained earlier in \cite{Calucci:1989hb}\cite{Calucci:1991sz}\cite{Accardi:2001ih}
in the tree-diagram approximation.
Remarkably, although dominant in the diagrammatic expansion of the
forward amplitude, as a consequence of unitarity and of the cutting
rules, there is no effect of loops to the inclusive cross section.

Notice also that, upon integration over all $p$, one obtains from (38) the
total multiplicity of jets from nucleus A multiplied by the cross-section
at a given impact parameter (the latter quantity being very close
to unity for a heavy nucleus). The result coincides with the average number of
{\it wounded partons} of the nucleus
A \cite{Calucci:1989hb}\cite{Calucci:1991sz}\cite{Accardi:2001ih}\cite{Bialas:1976ed}\cite{Accardi:2000ry}. From (38) we find
\beq \frac {d\sigma}{d^2\beta dy}= \int
d^2bT_A(b-\beta)P_A(x)\Big\{1-e^{-T_B(b)F_B(x,0)}\Big\} \eeq


\section{Numerical results}
The distribution in rapidity and transverse momentum, Eq.(38), is obtained form the
elementary inclusive cross-section $I(u,w,;p)$
of two partons colliding
with fractional momenta $u$ and $w$. For the gluon- gluon scattering one has
\cite{Eskola:1988yh}\cite{Eskola:2001rx}
\beq
\frac{d\sigma}{dy d^2p}=\frac{9\alpha_s^2}{2p^4}
\int dy'dudw uG(u) wG(w)
\left(1-\frac{p^2}
{suw}\right)^3\delta\left(u-x-\frac{p^2}{sx'}\right)
\delta\left(w-x'-\frac{p^2}{sx}\right)
\eeq
with
\[ x=\frac{p}{\sqrt{s}}e^y,\ \ \  x'=\frac{p}{\sqrt{s}}e^{-y'}\]
the fractional momenta of the observed and recoiling gluons
and $G(u)$, $G(w)$ the distributions of the initial gluons, which we assume to
have a small transverse momentum component. Eq. (40) exhibits a  kinematical
constraint between the
longitudinal and transverse momenta of the final gluons. In particular their
scaling variables are somewhat lower than those of the initial gluons.
Rigorously speaking this goes beyond the Glauber approximation,
which, as stated above, assumes the conservation of longitudinal momenta.
At large $s$ however the problem is felt only close to the limiting value
$p^2\sim p_{max}^2=suw/4$. For $p$ much smaller
than its limiting value, at $y'<0$ one can take $G(u)$ out of the
integral (40), at a value of momentum fraction corresponding to the momentum
of the observed gluon, $u=x$. The rest of the integral
can be transformed to the integration variable $w$. In the region $y'>0$
one can similarly take $G(w)$ out of the integral
and transform the integration to the variable $u$. In this way one obtains the
final cross-section in a quasi-Glauber form (see Appendix 1. for the details).
\beq
\frac{d\sigma}{dy d^2p}=\frac{9\alpha_s^2}{p^4}\Big[xG(x)
\int_{w_{min}} \frac{dw}{w}wG(w)
\left(1-\frac{p^2}
{sxw}\right)^2 +\Big(y\to -y\Big)\Big]
\eeq
where
\[w_{min}=\frac{p}{\sqrt{s}}(1+e^{-y})\]
We have checked that the approximations made in reducing (40) to (41) at
energies above 200 GeV
change the spectrum by no more than 10\% for $p<40$ GeV/c.
The two terms in (41) have evidently the meaning of the contributions from
the projectile and target partons.
At $y=0$ they are naturally equal.
To take into account the contribution of quarks, following
\cite{Eskola:1988yh}\cite{Eskola:2001rx}\cite{Pancheri:qg},
we introduce an effective parton density $P(x)$ and make the
substitution
\beq
xG(x)\to xP(x)=xG(x)+\frac{4}{9}[xQ(x)+x\bar{Q}(x)]
\eeq
Finally, to fit the existing experimental data, we multiply the cross-section by a
$K$-factor. With $K=3$ and the partonic
densities taken from \cite{Gluck:1998xa} we obtain a rather satisfactory
agreement with the experimental data on jet production on protons at 200 GeV
\cite{Albajar:1988tt}, as illustrated in Fig. 2.

The actual expression of the elementary inclusive cross section entering in Eq.(33) is
\beq
I(x,w;p)=(2\pi)^2K\frac{9\alpha_s^2}{p^4}\left(1-\frac{p^2}
{sxw}\right)^2
\theta\Big(sxw-p^2(1+e^y)\Big)
\eeq
Of course the expression holds only for sufficiently high values
of the transferred momenta. So in the integral (33) we standardly
restrict the integration  to $p>p_0$ where
$p_0$ is the infrared cutoff parameter.
We neglect the change in the initial nuclear parton distributions as
compared to the protonic ones (the EMC effect) and take $P_A(x)=AP(x)$
where $P(x)$ is the effective partonic distribution in the proton (42).
In our calculations we use the partonic densities of \cite{Gluck:1998xa} at the leading order.

We restrict ourselves to the production of jets at center rapidity $y=0$.
To simplify the calculations we also consider central collisions ($\beta=0$)
of identical nuclei and take the nuclei as  spheres of radius $R_A$ with a
constant density, so that
\beq
T_A(b)=\frac{2}{V_A}\sqrt{R_A^2-b^2}
\eeq
Then in (38) the integration  over $b$ can be done explicitly and we get
\beq
(2\pi)^2\frac {d\sigma}{d^2\beta dy d^2p}=2xAP(x)
\int d^2re^{ipr}
\Big(Z(z)-Z(z_0)\Big)
\eeq
where
\beq
Z(z)=e^z\left(\frac{6}{z^3}-\frac{6}{z^2}+\frac{3}{z}\right)-1-\frac{6}{z^3}
\eeq
\beq
z(r)=\frac{3A}{2\pi R_A^2}(f(r)-f(0)),\ \ z_0=-\frac{3A}{2\pi R_A^2}f(0))
\eeq
and
\beq
f(r)=
\frac{9}{2} K\int d^2p' e^{ip'r}\theta ({p'}^2-p_0^2)
\int_{w_{min}} \frac{dw}{w}wP(w,{p'}^2)\frac{\alpha^2_s({p'}^2)}{{p'}^4}
\left(1-\frac{{p'}^2}
{sxw}\right)^2
\eeq
In (48) we have chosen the scales for $\alpha_s$ and $P(w)$
to be equal to the transverse momentum squared of the recoil parton (i.e.
to the square of the transverse momentum transferred to the observed jet).
The factor 2 in (45) takes into account jets produced by projectile and target
partons. For the running coupling constant we have taken
$\Lambda_{QCD}=0.3$ GeV/c with four flavours.

Some care is needed in determining the scale of the
initial partonic distribution $P(x)$. It is related to the momentum
transferred in the first collision. In the case of a single collision
the scale is of order $p^2$, the transverse momentum
squared of the observed jet. In case of multiple collisions the
scale may however be substantially smaller. In our calculations we separated the
single scattering contribution from the general expression (45), where we have taken
$P(x)$ at the scale $p^2$. In the remaining part, corresponding to double
and higher scatterings, we have taken $P(x)$ at the scale $p_0^2$.

Even with the above simplifications the calculation of the final inclusive
distribution is far from trivial since it involves direct and inverse
Bessel transforms. We were able to obtain more or less stable results only for
values of $p$ not higher than $\sim$20 GeV/c, where,
fortunately, we
expect hard rescatterings to be most relevant.

We have considered central Pb-Pb collisons at 200 and 6000 GeV, corresponding to
RHIC and LHC energies.
We have taken $p_0=2$ GeV/c in accordance with
the analysis of  particle production data \cite{Albajar:1988tt}.
As for the K-factor we have considered values in the range 1 $\div$ 3.
As mentioned, at 200 GeV the value  $K=3$ is favoured by the proton data \cite{Gluck:1998xa}.
The analysis in \cite{Eskola:2002kv} found that while $p_0$ does not
practically change with energy, the K-factor clearly diminishes
to values close to unity.

In Figs. 3 and 4 we present our results in the form of  ratios of the
distribution (45), with all rescatterings included, to the contribution of
the single scattering only (optical approximation). The latter is obviously
just the proton-proton distribution multiplied by a nuclear factor
\beq
F_A=\frac{9A^2}{8\pi R_A^2}
\eeq

\section {Conclusions}

Given the very large energies in heavy ion collision at the LHC, a common
expectation is that
global features of the typical event will be within reach of
a perturbative QCD approach. It should nevertheless be emphasized the such a possibility is
far from trivial, the capability of
making statements on properties of an average inelastic event implying
qualitative improvements of present
understanding of strong interaction dynamics. In a typical heavy ion collision event
at the LHC several different mechanisms, originated by the complexity of the
interactions states, may play important roles giving rise to
a rather structured interaction, which needs to be brought to light to gain
the capability of making quantitative statements.

While an exhaustive description of the whole interaction process, even if limited
to the hard component only, is still out of reach, various
features which are likely to play an important role are presently under extensive investigation
in the
literature\cite{Baier:1998kq}\cite{Wiedemann:2000za}\cite{Wiedemann:2000tf}\cite{Gyulassy:2000fs}.
The topic which we have addressed in the present paper
is that of multiparton interactions, where several
partons in the initial state are linked in a hard collision process, which is a
rather natural mechanism to consider in a very dense interacting system.
One of the reasons of interest is that such a process
is able to restore the local isotropy in transverse space of the
black disk limit of the interaction,
which is broken by a two parton initiated process at the lowest
orders in $\alpha_S$\cite{Accardi:2000ry}.

Previous attempts to face this issue were based on a purely probabilistic description
of the hard component of the multiparton interaction, which has a solid
support for disconnected multiparton processes, while it has not such a
strong basis in the case of a connected multiparton collision. Within an incoherent
approach, interference effects
are in fact neglected altogether, which, in particular, implies disregarding all
off-diagonal contributions, namely ignoring the well known problem of loops in nucleus-nucleus
collisions, which give, on the contrary, the dominant contribution to the overall interaction
amplitude, as discussed in the first part of this paper. To approach
the problem we have studied the simplest case where only elastic partonic interactions
are taken into account and the overall nucleus-nucleus
$S$-matrix is factorized as a product of elementary parton-parton $S$-matrices.
In our approach production processes at the partonic level are therefore completely disregarded.
Since the complexity of the interaction is described by
the Glauber prescription of factorization of the $S$-matrix, any
amplitude for a $n$-parton interaction process is represented by a convolution
of two-partons
interaction amplitudes, so that the parton-nucleus amplitude
is expressed by a series of on shell rescattering terms.
Notice that, as in the case of the canonical Glauber approach to
hadron-nucleus collisions, this by no means implies a space-time ordering between different
interactions. It only implies that a connected $n$-body interaction process is well approximated
by a product of two-body interactions, which basically means that, in a dispersive
representation of the projectile-exchanged gluon amplitude,
the pole contribution is dominant. As for the quantitative relevance
of hard rescatterings, our numerical study confirms that, at transverse momenta
of $\sim$ 20 GeV, the
effects are large both at RHIC and LHC energies.

The feature which, in our opinion, makes this approach interesting,
even if some non-secondary features, as energy loss, are not included in the picture
of the interaction,
is that it allows one to obtain,
without approximations, the inclusive transverse spectrum of minijets,
which is therefore an exact consequence of the
$S$-matrix considered.
This feature allows one to argue that the result
might represent a good starting point for the ambitious
program of accomplishing an exhaustive description
of hard spectra in nucleus-nucleus collisions.
\vskip.25in
{\bf Acknowledgment}
\vskip.15in
This work was partially supported by the RFFI Grant 01-02-17137 of Russia
and Italian Ministry of University and of
Scientific and Technological Researches (MURST) by the Grant COFIN2001.
\vskip.25in

\section {Appendix 1. Derivation of (41)}
Let us split the integration over $y'$ in (40) in two parts, $y'<0$ and
$y'>0$. Consider the part with $y'<0$. It   corresponds to
observed gluons coming from the projectile. Integrating over $u$ we get for this
part
\beq
\left(\frac{d\sigma}{dy d^2p}\right)_{y'<0}=\frac{9\alpha_s^2}{p^4}
\int_{p/\sqrt{s}}\frac{dx'}{x'}\int dw uG(u)wG(w)
\left(1-\frac{p^2}{suw}\right)^3
\delta\left(w-x'-\frac{p^2}{sx}\right)
\eeq
with $u=x+p^2/(sx')$. Further integration over $x'$ gives
\beq
\left(\frac{d\sigma}{dy d^2p}\right)_{y'<0}=\frac{9\alpha_s^2}{p^4}
\int_{w_{min}}\frac{dw}{w} uG(u)wG(w)\frac{w}{w-p^2/(sx)}
\left(1-\frac{p^2}{suw}\right)^3
\eeq
where now
\beq
u=x\left(1+\frac{xp^2}{sw-p^2}\right)
\eeq
and $w_{min}$ is determined from the condition that $x'>p/\sqrt{s}$
which gives
\beq
w_{min}=\frac{p}{\sqrt{s}}+\frac{p^2}{xs}
\eeq

Now we make our crucial approximation to neglect the second term in (53)
and put $u=x$. This leads directly to the first term in (41).

The contribution from the region $y'>0$ is transformed in a similar manner.
Now we integrate over $w$ to get
\beq
\left(\frac{d\sigma}{dy d^2p}\right)_{y'>0}=\frac{9\alpha_s^2}{p^4}
\int^{p/\sqrt{s}}\frac{dx'}{x'}\int dw uG(u)wG(w)
\left(1-\frac{p^2}{suw}\right)^3
\delta\left(u-x-\frac{p^2}{sx'}\right)
\eeq
with $w=x'+p^2/(sx)$. Further integration over $x'$ gives
\beq
\left(\frac{d\sigma}{dy d^2p}\right)_{y'<0}=\frac{9\alpha_s^2}{p^4}
\int_{u_{min}}\frac{du}{u} uG(u)wG(w)\frac{u}{u-x}
\left(1-\frac{p^2}{suw}\right)**3
\eeq
Here
\beq
w=\frac{p^2}{sx}\left(1+\frac{x}{u-x}\right)
\eeq
and $u_{min}$ is determined from the condition $x'<p/\sqrt{s}$ to be
\beq
u_{min}=x+\frac{p}{\sqrt{s}}
\eeq
Again our approximation consists in neglecting the second term in (56). After
that we obtain the second term in (41).

\section{Appendix 2. Fits to the existing jet data [7]}
The fits have the form
\beq
\frac{d\sigma}{dy dp}= \frac{A}{(p+p_0)^a},\ \frac{\mu{\rm b}}{\rm GeV/c}
\eeq
Here follows the table of the parameters with the corresponding values of
$\chi^2$

\begin{center}
\begin {tabular}{|l|r|r|r|}
\hline
 &200 GeV & 500 GeV & 900 GeV\\\hline
A & (0.977$\pm$0.047)E+10 & (1.00$\pm$ 0.97)E+10 &(1.00$\pm$ 0.98)E+10\\
a & 8.028$\pm$ 0.095      & 7.119$\pm$ 0.033     & 6.433$\pm$ 0.034    \\
$p_0$   & 4.14$\pm$ 0.32  & 5.83$\pm$ 0.22     & 8.50$\pm$ 0.28    \\
$\chi^2$& 3.84            & 5.514      & 16.4      \\\hline
A & (0.91$\pm$ 0.48)E+8  & (0.175$\pm$ 0.045)E+8 &(0.379$\pm$ 0.065)E+7\\
a & 6.68$\pm$ 0.23      & 5.383$\pm$ 0.096     & 4.402$\pm$ 0.068    \\
$p_0$   & 2.0 (fixed) & 2.0 (fixed)& 2.0 (fixed)\\
$\chi^2$& 4.54        & 10.87      & 33.6       \\\hline
\end{tabular}
\end{center}

\section{Appendix 3. Details of the numerical calculation}
Integration over the angles transforms (45) and (48) in
\beq
\frac {d\sigma}{d^2\beta dy d^2p}=\frac{1}{\pi} xAP(x)
\int rdr{\rm J}_0(pr)
\Big(Z(z)-Z(z_0)\Big)
\eeq
and
\beq
f(r)=
9\pi K\int_{p_0} p'dp'{\rm J}_0(p'r)\frac{\alpha^2_s({p'}^2)}{{p'}^4}
\int_{w_{min}} \frac{dw}{w}wP(w,{p'}^2)
\left(1-\frac{{p'}^2}{sxw}\right)^2
\eeq
where according to (40) and (41) at $y=0$ $x=p/\sqrt{s}$ and
\beq
w_{min}=\frac{2{p'}^2}{sx}=\frac{2{p'}^2}{p\sqrt{s}}
\eeq
One observes that the two integrations in (60) over $w$ y $p'$ do not decouple,
which makes the calculation rather complicated.

\section{Figure captions}

1. Diagrams illustrating the interactions of two partons from the projectile with two partons from the target.

2. Jet distributions in $p\bar p$ collisions at 200 GeV at center rapidity.
The lower curve is a fit to the experimental data [7]. The upper curve shows predictions from Eq. (41) with $K=3.0$.

3. Ratios of the total distributions to the single scattering contributions
(optical approximation) for central Pb-Pb collisions at 200 GeV and center
rapidity
calculated from(45) with $p_0=2$ GeV/c.

4. Ratios of the total distributions to the single scattering contributions
(optical approximation) for central Pb-Pb collisions at 6000 GeV and center
rapidity calculated from(45) at $y=0$ with $p_0=2$ GeV/c.


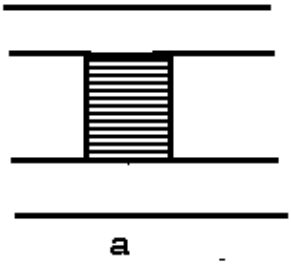
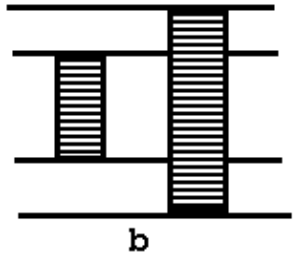
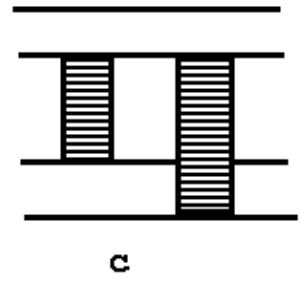
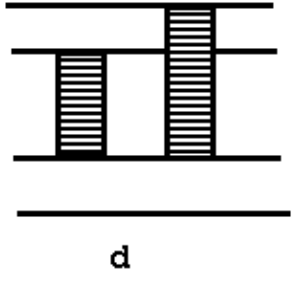
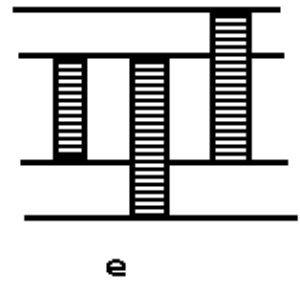
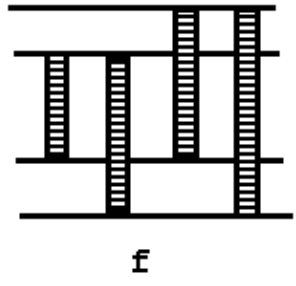
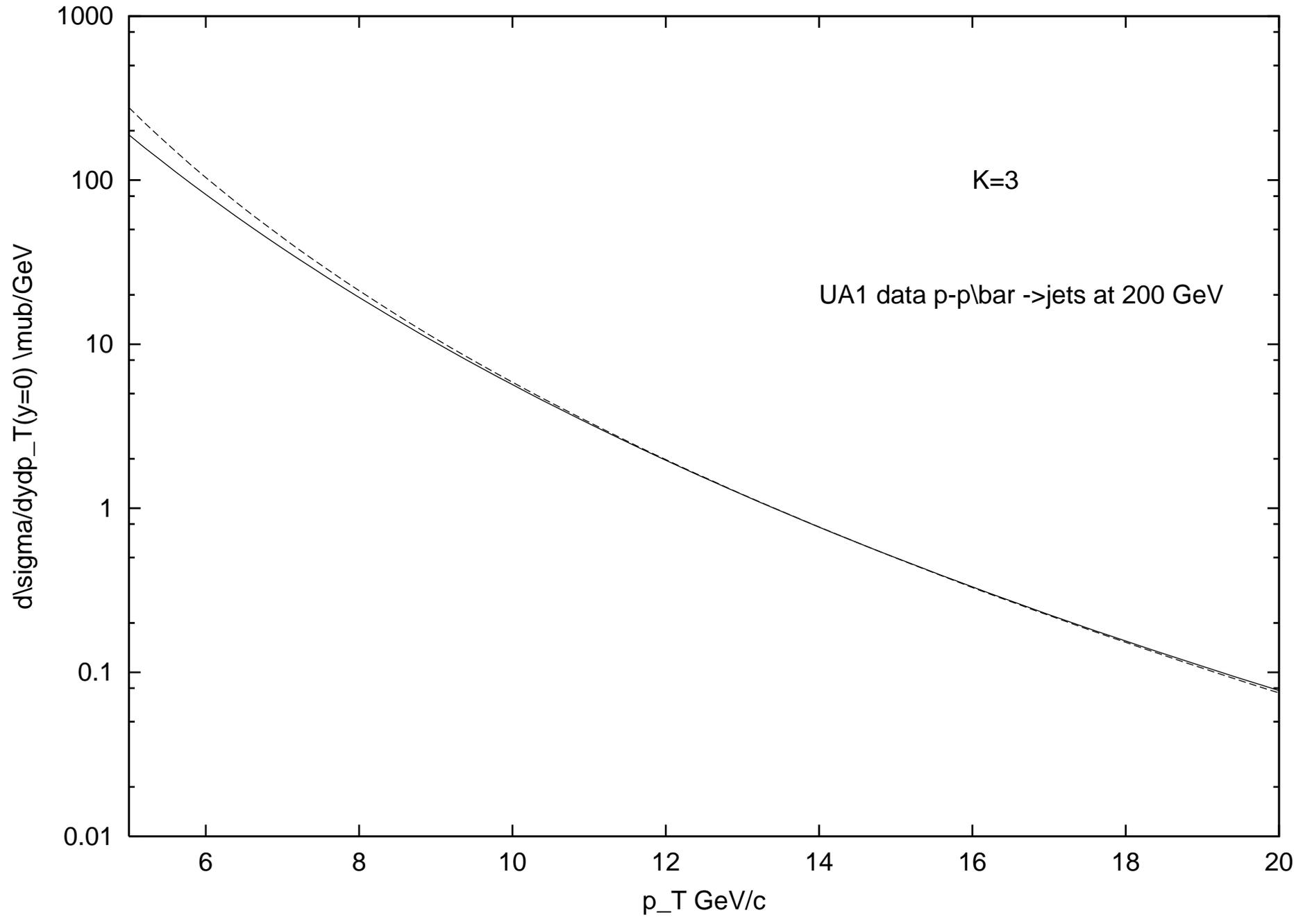
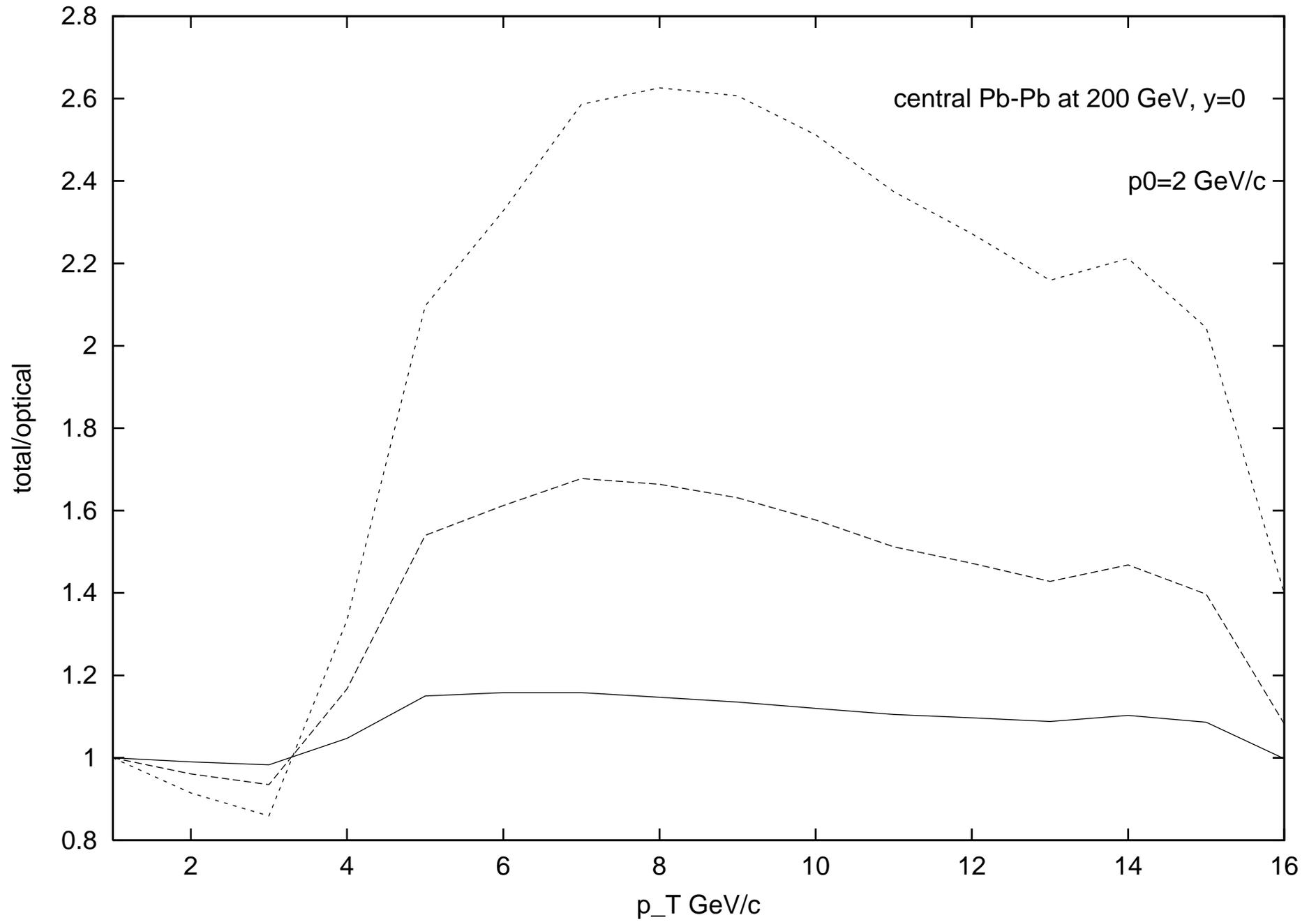
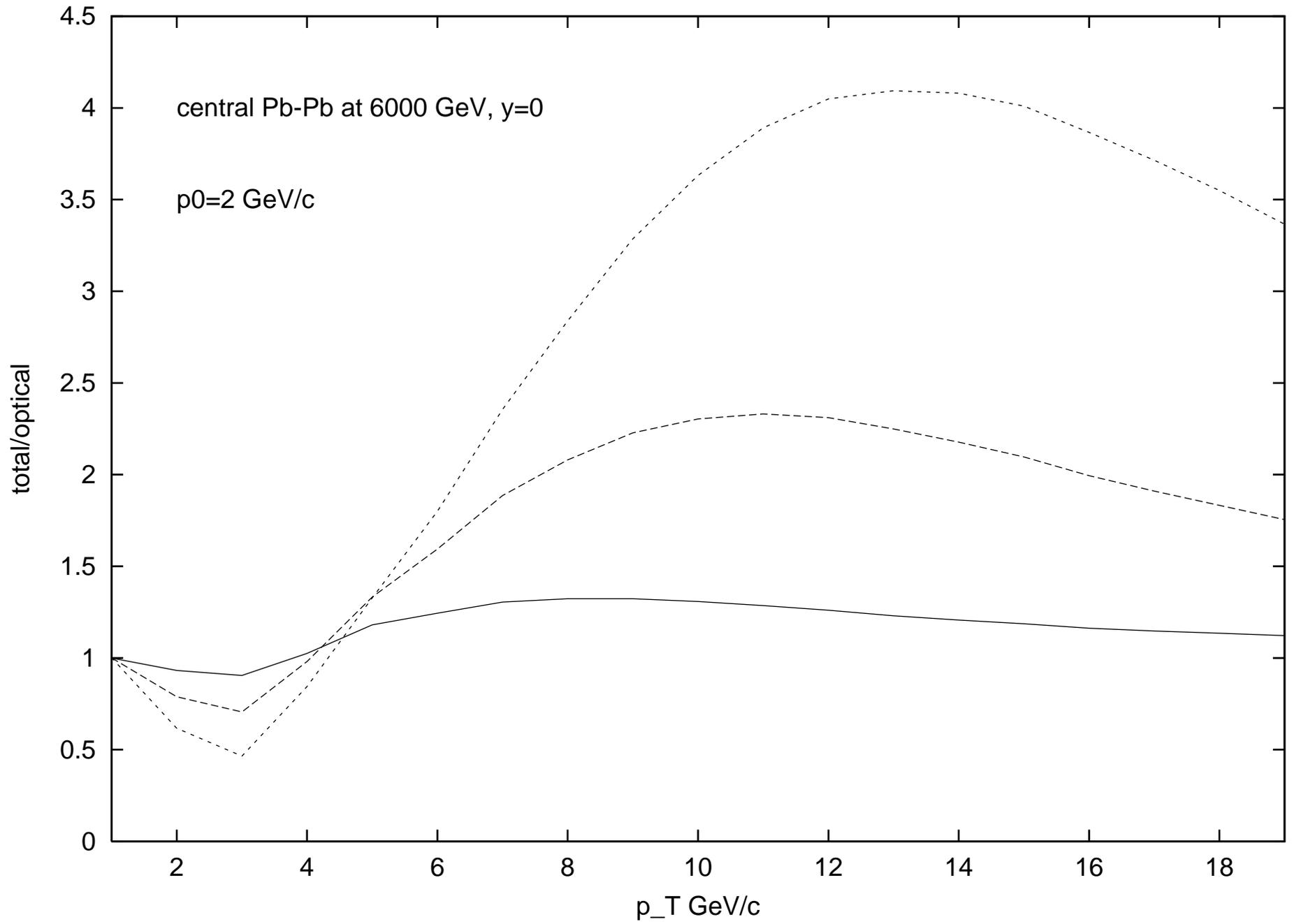
\end{document}